\documentclass[nofootinbib,floatfix,twocolumn,superscriptaddress,a4paper,prl]{revtex4}

\usepackage{amsmath}
\usepackage{amssymb}
\usepackage{graphicx}
\usepackage{color}
\usepackage[T1]{fontenc} 
\usepackage{slashed}

\renewcommand{\arraystretch}{1.6}
\newcommand{\hc}{\mathrm{h.c.}}

\newcommand{\nn}{\nonumber}
\newcommand{\id}{\mathbb{I}}
\newcommand{\npar}{\texttt{\#}}

\newcommand{\AddrIFIC}{%
Instituto de F\'{\i}sica Corpuscular (CSIC-Universitat de Val\`{e}ncia),
Apdo. 22085, E-46071 Valencia, Spain
}

\newcommand{\AddrUV}{%
Departamento de Matem\'aticas, Universitat de Val\`encia, E-46100 Burjassot, 
Val\`encia, Spain}

\preprint{IFIC/18-46}
\begin{document}

\title{Master Majorana neutrino mass parametrization}

\author{I. Cordero-Carri\'on}\email{isabel.cordero@uv.es}
\affiliation{\AddrUV}

\author{M. Hirsch} \email{mahirsch@ific.uv.es }
\affiliation{\AddrIFIC}

\author{A. Vicente} \email{avelino.vicente@ific.uv.es}
\affiliation{\AddrIFIC}

\begin{abstract}
After introducing a master formula for the Majorana neutrino mass
matrix we present a master parametrization for the Yukawa matrices
automatically in agreement with neutrino oscillation data. This
parametrization can be used for any model that induces Majorana
neutrino masses. The application of the master parametrization is also
illustrated in an example model, with special focus on its lepton
flavor violating phenomenology.
\end{abstract}
\maketitle

{\em Introduction:} The Standard Model (SM) of particle physics stands
as one of the most successful physical theories ever built. However,
despite its tremendous success, it cannot describe all particle
physics phenomena. Neutrino oscillation experiments have firmly
established that neutrinos have non-zero masses and mixings, hence
demanding an extension of the SM that accounts for them. 

Many neutrino mass models have been proposed. A short list, to quote
only a few reviews and general classification papers, includes models
with Dirac \cite{Ma:2016mwh,CentellesChulia:2018gwr} or Majorana
neutrinos \cite{Ma:1998dn}, with neutrino masses induced at tree-level
or radiatively at 1-loop/2-loop \cite{Cai:2017jrq} or 3-loop
\cite{Cepedello:2018rfh}, at low- \cite{Boucenna:2014zba} or
high-energy scales, and by operators of dimension 5 or higher
dimensionalities \cite{Anamiati:2018cuq}. 

The goal of this letter is twofold. First, we will introduce a master
formula that unifies all Majorana neutrino mass models, which can be
regarded as particular cases of this general expression. And second,
we will present a master parametrization for the Yukawa matrices
appearing in this formula. The parametrization presented in this
letter extends previous results in the literature \cite{Casas:2001sr}
and can be used for any model that induces Majorana neutrino
masses. \\

{\em The master formula:} With full generality, a Majorana neutrino
mass matrix can be written in the form
\begin{equation} \label{eq:master}
m = f \, \left( y_1^T \, M \, y_2 + y_2^T \, M^T \, y_1 \right) \, .
\end{equation}
Here $m$ is the $3 \times 3$ complex symmetric neutrino mass matrix,
\footnote{We focus on the case of 3 generations, because there are
  only three active neutrinos. It is straightforward to generalize to
  a larger number, if one wants to include for example light sterile
  neutrinos.}  which can be diagonalized as
\begin{equation} \label{eq:m-U}
D_m = \text{diag} \left( m_1, m_2, m_3 \right) = U^T \, m \, U \, ,
\end{equation}
with $U$ a $3 \times 3$ unitary matrix ($U^\dagger U = U U^\dagger =
\id_3$). The matrices $y_1$ and $y_2$ are general dimensionless $n_1
\times 3$ and $n_2 \times 3$ complex matrices, respectively, and $M$
is a $n_1 \times n_2$ complex matrix with dimension of mass. Without
loss of generality, we will assume $n_1 \geq n_2$. We note that $m$
must contain at least two non-vanishing eigenvalues in order to
explain neutrino oscillation data. Therefore, in the following we
consider $r_m = \text{rank}(m) = 2 \, \text{or} \, 3$.

Eq. \eqref{eq:master} is a \textit{master formula} valid for
\textit{all} Majorana neutrino mass models. This can be illustrated
with several examples. The simplest one is based on the popular seesaw
mechanism
\cite{Minkowski:1977sc,Yanagida:1979as,Mohapatra:1979ia,GellMann:1980vs,Magg:1980ut,Schechter:1980gr},
in particular on the standard type-I seesaw with $3$ generations of
right-handed neutrinos. The light neutrino mass matrix in this model
is given by $m = - \langle H^0 \rangle^2 \, y^T M_R^{-1} y$, an
expression that can be obtained with the master formula by taking
$f=-1$, $y_1 = y_2 = y/\sqrt{2}$ and $M = \langle H^0 \rangle^2 \,
M_R^{-1}$. Here $\langle H^0 \rangle = v / \sqrt{2}$ is the SM Higgs
vacuum expectation value (VEV) and $M_R$ the Majorana mass matrix for
the right-handed neutrinos. Moreover, these matrices are all $3 \times
3$ and hence $n_1 = n_2 = 3$ in this model. The mass matrices of more
complicated Majorana neutrino models can also be accommodated with the
master formula. For instance, the inverse seesaw
\cite{Mohapatra:1986bd} would correspond to $M = \langle H^0 \rangle^2
\, (M_R^T)^{-1} \mu M_R^{-1}$, with $\mu$ the small lepton number
violating mass scale in this model, whereas the scotogenic model
\cite{Ma:2006km}, in which neutrino masses are induced at the 1-loop
level, corresponds to $f = \lambda_5 / (16 \pi^2)$ and $M = \langle
H^0 \rangle^2 \, M_R^{-1} F_{\rm loop}$, with $\lambda_5$ the coupling
of the quartic term $(H^\dagger \eta)^2$ involving the standard ($H$)
and inert ($\eta$) scalar doublets, and $F_{\rm loop}$ a matrix of
loop functions. In particular, models with $y_1 \ne y_2$ can be
described with the master formula, as shown below with the specific
example of the BNT model \cite{Babu:2009aq}. \\

{\em The master parametrization:} Our goal after introducing the
master formula in Eq. \eqref{eq:master} is to establish a
parametrization of the $y_1$ and $y_2$ Yukawa matrices with three
properties:
\begin{itemize}
\item {\bf General:} valid for all models.
\item {\bf Complete:} containing all the degrees of freedom in the model.
\item {\bf Programmable:} easy to use in phenomenological analyses.
\end{itemize}

{
\renewcommand{\arraystretch}{1.4}
\begin{table*}[t]
\centering
{\setlength{\tabcolsep}{2em}
\begin{tabular}{| c c c c |}
\hline  
Matrix & Dimensions & Property & Real parameters \\
\hline
\hline    
$X_1$ & $(n_2-n) \times 3$ & Absent if $n = n_2$ & $6(n_2-n)$ \\
$X_2$ & $(n_1-n_2) \times 3$ & Absent if $n_1 = n_2$ & $6(n_1-n_2)$ \\
$X_3$ & $(n_2-n) \times 3$ & Absent if $n = n_2$ & $6(n_2-n)$ \\
$W$ & $n \times r$ &  & $r (2n-r)$ \\
$T$ & $r \times r$ & Upper triangular with $(T)_{ii} > 0$ & $r^2$ \\
$K$ & $r \times r$ & Antisymmetric & $r (r-1)$ \\
$\bar B$ & $(n-r) \times 3$ & Absent if $n = r$ & $6(n-r)$ \\
$C_1$ & $r \times 3$ & Case-dependent & $0$ or $2$ \\
$C_2$ & $3 \times 3$ & Case-dependent & - \\
\hline
\end{tabular}
}
\caption{Matrices containing free parameters in the master
  parametrization. Even though the matrix $C_2$ does not contain any
  free parameter, we include it in this list since its form depends on
  the values of $r_m$ and $r$.
\label{tab:matrices}
}
\end{table*}
}

We will call this parametrization of the Yukawa matrices the
\textit{master parametrization}. We now proceed to present it. The
Yukawa matrices $y_1$ and $y_2$ can be generally parametrized as
\begin{align}
y_1 & = \frac{1}{\sqrt{2 \, f}} \, V_1^\dagger \, \left( \begin{array}{c}
\Sigma^{-1/2} \, W \, A \\ X_1 \\ X_2 \end{array} \right) \, 
\bar{D}_{\sqrt{m}} \, U^\dagger \, , \label{eq:par1} \\
y_2 & = \frac{1}{\sqrt{2 \, f}} \, V_2^\dagger \, \left( \begin{array}{c}
\Sigma^{-1/2} \, \widehat W^\ast \, \widehat B \\ X_3 \end{array} \right) 
\, \bar{D}_{\sqrt{m}} \, U^\dagger \, . \label{eq:par2}
\end{align}
Several matrices have been defined in the previous two expressions,
where $\ast$ denotes the conjugate matrix. We have defined the matrix
$\bar{D}_{\sqrt{m}}$ as
diag$\left(\sqrt{m_1},\sqrt{m_2},\sqrt{m_3}\right)$ if $r_m=3$ or
diag$\left(\sqrt{m_1},\sqrt{m_2},\sqrt{v}\right)$ if $r_m=2$. In fact,
$v$ can be replaced in this definition by any non-vanishing reference
mass scale since it is a dummy variable that drops out in the
calculation of the neutrino mass matrix. A singular-value
decomposition has been applied to the matrix $M$,
\begin{equation} \label{eq:M-SVD}
M = V_1^T \, \widehat \Sigma \, V_2 \, ,
\end{equation}
where $\widehat \Sigma$ is a $n_1 \times n_2$ matrix that can be written as
\begin{equation}
\widehat \Sigma = \left(\begin{array}{c} \begin{array}{cc} 
\Sigma & 0 \\ 0 & 0_{n_2-n} \end{array} \\ \hline 0_{n_1-n_2} 
\end{array}\right) \, ,
\end{equation}
and $\Sigma =
\text{diag}\left(\sigma_1,\sigma_2,\dots,\sigma_n\right)$ is a
diagonal $n \times n$ matrix containing the positive and real singular
values of $M$ ($\sigma_i>0$). Therefore, we define $n$ as the number
of non-zero singular values of the matrix $M$. Since the total number
of singular values of $M$ is $n_2$, it is clear that $n \leq n_2$. It
is possible to have vanishing singular values which are specifically
encoded in the zero square $(n_2-n)\times(n_2-n)$ matrix
$0_{n_2-n}$. $V_1$ and $V_2$ are $n_1 \times n_1$ and $n_2 \times n_2$
unitary matrices and can be found by diagonalizing the square matrices
$M M^\dagger$ and $M^\dagger M$, respectively. $X_1$, $X_2$ and $X_3$
are, respectively, $(n_2-n)\times 3$, $(n_1-n_2)\times 3$ and
$(n_2-n)\times 3$ arbitrary complex matrices with dimensions of
mass$^{-1/2}$. $\widehat W$ is an $n\times n$ matrix defined as
\begin{equation}
\widehat W = \left(W \quad \bar{W}\right) \, ,
\end{equation}
where $W$ is an $n\times r$ complex matrix, with $r=\text{rank}(W)$,
such that $W^\dagger W = W^T W^* = \id_r$, and $\bar{W}$ is an
$n\times(n-r)$ complex matrix, built with vectors that complete those
in $W$ to form an orthonormal basis of $\mathbb{C}^n$. Therefore,
$\widehat W$ is a unitary complex $n\times n$ matrix. $A$ is an
$r\times 3$ matrix, which can be written as
\begin{equation}
A = T \, C_1 \, ,
\end{equation}
with $T$ an upper-triangular $r\times r$ invertible square matrix
with positive real values in the diagonal, and $C_1$ is an $r\times 3$
matrix. $\widehat B$ is an $n\times 3$ complex matrix defined as
\begin{equation}
\widehat B = \left( \begin{array}{c} B \\ \bar{B} \\ \end{array} \right) \, ,
\end{equation}
with $\bar{B}$ an arbitrary $(n-r)\times 3$ complex matrix and $B$ an
$r\times 3$ complex matrix given by
\begin{equation} \label{eq:Bexp}
B \equiv B\left( T , K , C_1 , C_2 \right) 
= \left( T^T \right)^{-1} \, \left[ C_1 \, C_2 + K \, C_1 \right] \, ,
\end{equation}
where we have introduced the antisymmetric $r\times r$ square matrix
$K$ and the $3\times 3$ matrix $C_2$. The exact form of the matrices
$C_1$ and $C_2$ depends on the values of $r_m$ and $r$. For $r_m = r =
3$ these matrices take the form
\begin{align}
C_1 = \id_3, \quad
C_2 = \id_3 + K_{12} \, \frac{T_{13}}{T_{11}} \, \left( \begin{array}{ccc}
0 & 0 & 0 \\
0 & 0 & 1 \\
0 & -1 & 0 \end{array} \right) \, , \label{eq:C1C2}
\end{align}
while the expressions for other cases, as well as a rigorous
mathematical proof of the master parametrization, will be given
elsewhere~\cite{future}. We summarize the matrices that appear in the
master parametrization and count their free parameters in
Tab. \ref{tab:matrices}. \\

{\em Parameter counting:} In order to guarantee that the master
parametrization is complete, a detailed parameter counting must be
performed. In full generality, one can write
\begin{equation} \label{eq:counting}
\npar_{\rm free} = \npar_{y_1} + \npar_{y_2} - \npar_{\rm eqs} - \npar_{\rm extra}
= 6 (n_1 + n_2) - \npar_{\rm eqs} - \npar_{\rm extra} \, ,
\end{equation}
where $\npar_{y_1} = 2 \cdot 3 \cdot n_1$ and $\npar_{y_2} = 2 \cdot 3
\cdot n_2$ are the number of real degrees of freedom in $y_1$ and
$y_2$, respectively, and $\npar_{\rm eqs}$ is the number of
independent (real) equations contained in Eq. \eqref{eq:master}. Since
this matrix equation is symmetric, one would naively expect to have 6
complex equations, which would then translate into 12 real
restrictions on the elements of $y_1$ and $y_2$. However, one can
check by direct computation that for $r=1$ one of the complex
equations is actually redundant and can be derived from the other
five. Therefore,
\begin{equation}
\npar_{\rm eqs} = \left\{ \begin{array}{l}
12 \quad \text{for} \, r=3 \, \text{or} \, 2 ,\\
10 \quad \text{for} \, r=1 .
\end{array} \right.
\end{equation}
Note that the case $r=1$ is allowed only because \eqref{eq:master} 
contains two terms, each of which in principle can be of 
rank 1, as long as the rank of the sum of both terms is 2.
Finally, $\npar_{\rm extra}$ is the number of extra (real)
restrictions imposed on $y_1$ and $y_2$. In the most common case of
the standard type-I seesaw one has $\npar_{\rm extra} = 0$. However,
scenarios with additional restrictions have $\npar_{\rm extra} \neq
0$. The total number of free parameters $\npar_{\rm free}$ must match the
sum of the number of free parameters in each of the matrices appearing
in the master parametrization of Eqs. \eqref{eq:par1} and
\eqref{eq:par2}. Therefore
\begin{align} 
\npar_{\rm free} &= \npar_{X_1} + \npar_{X_2} 
+ \npar_{X_3} + \npar_{A} + \npar_{W} 
+ \npar_{B} + \npar_{\bar{B}} + \npar_{C_1} 
\nonumber \\
&= \npar_{X_1} + \npar_{X_2} 
+ \npar_{X_3} + \npar_{T} + \npar_{W} 
+ \npar_{K} + \npar_{\bar{B}} + \npar_{C_1} \, . \label{eq:counting2}
\end{align}
In the previous expressions we have taken $\npar_{\bar{W}} = 0$ and
assigned all the free parameters in the product $\bar{W} \, \bar{B}$
to $\bar{B}$. This is possible because these two matrices always
appear in the combination $\bar{W} \, \bar{B}$ and, given that all the
parameters contained in $\bar{B}$ are free, $\npar_{\bar{W}\bar{B}}
\equiv \npar_{\bar{B}}$. 

It proves convenient to discuss a particular example in order to
understand the general parameter counting procedure. Let us consider
$n_1 = n_2 = n = 3$ and focus on a scenario with $(r_m,r) = (3,3)$. In
this case $\widehat \Sigma \equiv \Sigma$, $\npar_{\rm eqs} = 12$ and
$\npar_{\rm extra} = 0$. Therefore, from Eq.\eqref{eq:counting}, one
finds $\npar_{\rm free}^{(3,3)}=24$. Using now
Eq. \eqref{eq:counting2}, one finds
\begin{equation}
\npar_{\rm free}^{(3,3)} = 24 = \npar_W^{(3,3)} + \npar_A^{(3,3)} 
+ \npar_B^{(3,3)} + \npar_{C_1}^{(3,3)} 
= 15 + \npar_{W}^{(3,3)} \, ,
\end{equation}
where $\npar_{W}^{(3,3)} = 9$ is the number of real free parameters in
the matrix $W$ in the $(3,3)$ case. We point out that
$\npar_{W}^{(3,3)} = 9$ also follows from the fact that $W$ is a
unitary $3 \times 3$ matrix, which makes a good consistency check of
the parameter counting we just performed. In addition, we note that
$\npar_A^{(3,3)} = 9$ and $\npar_B^{(3,3)} = 6$. \\

{\em The Casas-Ibarra limit:} One must finally compare the master
parametrization to previously known parametrizations in the
literature. In particular, let us compare to the Casas-Ibarra
parametrization~\cite{Casas:2001sr}. As already explained above, the
type-I seesaw corresponds to $y_1 = y_2 = y/\sqrt{2}$, $n_1 = n_2 = n
= r = 3$, $f=-1$ and $M = \langle H^0 \rangle^2 M_R^{-1}$. Furthermore,
in this model the symmetric matrix $M$ can be diagonalized by a single
matrix, $V_1 = V_2$, which can be taken to be the identity if the
right-handed neutrinos are in their mass basis, and the matrices
$X_{1,2,3}$, $\overline W$ and $\overline B$ drop from all the
expressions. Finally, imposing $y_1 = y_2$ is equivalent to $W^T W A =
B$. Solving this matrix equation leads to $B = \left(A^T\right)^{-1}$
and allows one to define $R = W \, A$, with $R$ a general $3 \times 3$
orthogonal matrix. Replacing all these ingredients into
Eqs. \eqref{eq:par1} and \eqref{eq:par2} one finds
\begin{equation}
y = \sqrt{2} \, y_1 = \sqrt{2} \, y_2 = i \, \Sigma^{-1/2} \, R \, D_{\sqrt{m}} \, U^\dagger \, ,
\end{equation}
which is nothing but the Casas-Ibarra parametrization for the type-I
seesaw Yukawa matrices. We note that $R$ can be identified with the
usual Casas-Ibarra matrix~\cite{Casas:2001sr}. Imposing $y_1 = y_2$
leads to $18 \, (= 9 \cdot 2)$ real constraints, this is, $\npar_{\rm
  extra} = 18$. Therefore, direct application of the general counting
formula in Eq. \eqref{eq:counting} leads to $\npar_{\rm free} =
6$. These are the free real parameters contained in $R$ which can be
parametrized by means of $3$ complex angles. We conclude that the
Casas-Ibarra parametrization can be regarded as a particular case of
the general master parametrization. \\

{\em An application:} The full power of the master parametrization is
better illustrated with an application to the BNT model
\cite{Babu:2009aq}. In addition to the SM particles, the model
contains three copies of the vector-like fermions $\psi_{L,R}$
transforming as $({\bf 1}, {\bf 3}, -1)$ under the SM gauge group and
an exotic scalar $\Phi$ transforming as $({\bf 1}, {\bf 4}, 3/2)$. The
quantum numbers of the new particles in the BNT model are given in
Table~\ref{tab:BNT}.

{
\renewcommand{\arraystretch}{1.4}
\begin{table}
\centering
{\setlength{\tabcolsep}{0.5em}
\begin{tabular}{| c c c c c |}
\hline  
 & generations & $\mathrm{SU(3)}_c$ & $\mathrm{SU(2)}_L$ & $\mathrm{U(1)}_Y$ \\
\hline
\hline    
$\Phi$ & 1 & ${\bf 1}$ & ${\bf 4}$ & $3/2$ \\
\hline
\hline    
$\psi_{L,R}$ & 3 & ${\bf 1}$ & ${\bf 3}$ & $-1$ \\ 
\hline
\end{tabular}
}
\caption{New particles in the BNT model.}
\label{tab:BNT}
\end{table}
}

The Lagrangian of the model contains the following pieces relevant for
neutrino mass generation
\begin{align} \label{eq:YukBNT}
-\mathcal L &\supset y_\psi \, \overline{L} \, H \, \psi_R + y_{\bar \psi} \, \overline{L^c} \, \Phi \, \psi_L + M_\psi \overline \psi \, \psi + \hc \, ,
\end{align}
where we have omitted $\mathrm{SU(2)_L}$ and flavor indices to
simplify the notation. The scalar potential of the model is given by
\begin{align} \label{eq:PotBNT}
\mathcal V &= M_{H}^2 |H|^2 + M_{\Phi}^2 |\Phi|^2 + \frac{1}{2} \lambda_1 \, |H|^4 + \frac{1}{2} \lambda_2 \, |\Phi|^4 \nn \\
&+ \lambda_{3} \, \left( |H|^2 |\Phi|^2 \right)_{\bf 1} + \lambda_{4} \, \left(|H|^2 |\Phi|^2\right)_{\bf 3} + \lambda_\Phi \, \left[ H^3 \, \Phi^{\dagger} + \hc \right] \, .
\end{align}    
Here $H$ is the SM Higgs doublet. We note that there are two possible
$\mathrm{SU(2)_L}$ contractions of $|H|^2 |\Phi|^2$, corresponding to
the $\lambda_3$ and $\lambda_4$ quartic terms. All the couplings in
the scalar potential must be real, with the exception of
$\lambda_\Phi$, which can be complex. The introduction of
$\lambda_\Phi \ne 0$ precludes the introduction of a non-vanishing
lepton number for $\Phi$. In fact, one can easily see that lepton
number is broken in two units in the BNT model. Furthermore, this term
induces a non-zero VEV for the neutral component of $\Phi$, $\Phi^0$,
which is given by
\begin{equation}
\langle \Phi^0 \rangle = \frac{v_{\Phi}}{\sqrt{2}} = \frac{\lambda_\Phi v^3}{2 \sqrt{2} M_\Phi^2} \, .
\end{equation}

\begin{figure}[t]
\centering
\includegraphics[width=0.4\textwidth]{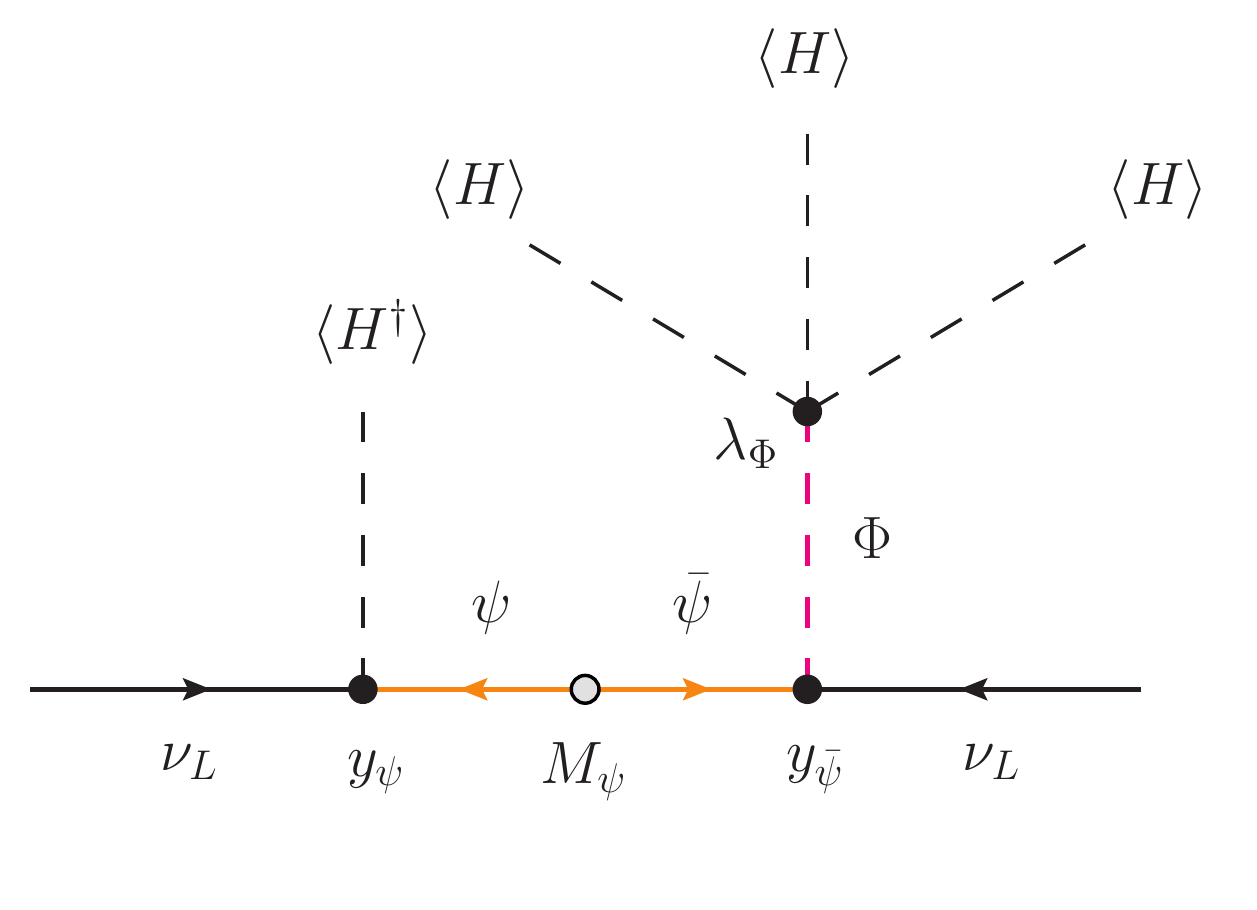}
\caption{Neutrino mass generation in the BNT model.} 
\label{fig:BNTmodel}
\end{figure}

In the BNT model, neutrino masses are generated at dimension 7 as
shown in Fig.~\ref{fig:BNTmodel}. The resulting expression for the
neutrino mass matrix is
\begin{align} \label{eq:mnuBNT}
m = \frac{\lambda_\Phi v^4}{4 M_\Phi^2} \, \left[ y_\psi^T \, M_\psi^{-1} \, y_{\bar \psi} + y_{\bar \psi}^T \, (M_\psi^{-1})^T \, y_{\psi} \right] \,.
\end{align}

The usual Casas-Ibarra parametrization cannot be applied in this model
since one has two independent $y_1 = y_\psi$ and $y_2 = y_{\bar \psi}$
Yukawa matrices. Therefore, in order to guarantee that the parameters
measured in neutrino oscillation experiments are correctly reproduced
one must make use of the master parametrization. In order to apply the
master parametrization we must first identify the different pieces
taking part of the neutrino mass expression in the BNT model,
Eq. \eqref{eq:mnuBNT}. By direct comparison to the master formula in
Eq. \eqref{eq:master} we identify
\begin{equation}
f = \frac{\lambda_\Phi v^2}{2 M_\Phi^2} \, , \quad M = \frac{v^2}{2} \, M_\psi^{-1} \, .
\end{equation}
Furthermore, in this model $y_1$, $y_2$ and $M$ are $3 \times 3$
matrices and then $n_1 = n_2 = 3$. One also has $3$ non-vanishing
singular values in $M$ and therefore $n = 3$ and $\widehat \Sigma
\equiv \Sigma$. Finally, taking $r = r_m = 3$, the matrices
$X_{1,2,3}$ and $\bar B$ are absent, while $C_1$ and $C_2$ are given
in Eq. \eqref{eq:C1C2}.

\begin{figure}[t]
 \centering
 \includegraphics[width=0.45\textwidth]{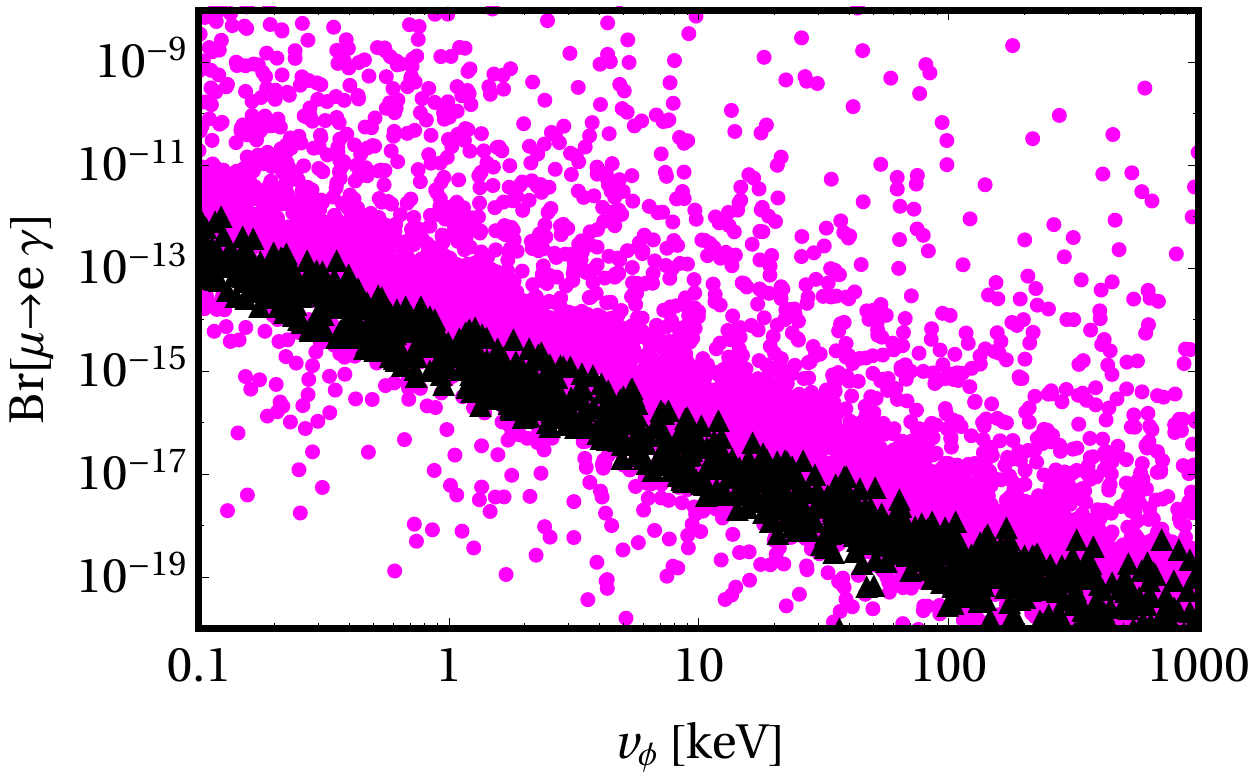}
 \caption{Example points for Br$(\mu\to e\gamma)$ in the BNT model as
   a function of the quadruplet VEV $v_{\Phi}$. This figure has been
   obtained by allowing the neutrino oscillation parameters to vary
   within the 3 $\sigma$ ranges determined by the global fit
   \cite{deSalas:2017kay}, assuming normal hierarchy, $M_{\Psi}$
   randomly taken in the interval $[0.5,2]$ TeV and $W = \id$. The
   purple points correspond to a scan in which the entries of the
   matrices $T$ and $K$ are randomly taken in the following ranges:
   $T_{ii} \in [0,2]$ and $K_{ij}$, $T_{ij}$ (with $i\ne j$) $\in
   [-1,1]$. The black points in the foreground correspond to a
   simplified scan with $T = \id$ and $K = 0$.
\label{fig:Ex1}}
\end{figure}

\begin{figure}[t]
 \centering
 \includegraphics[width=0.45\textwidth]{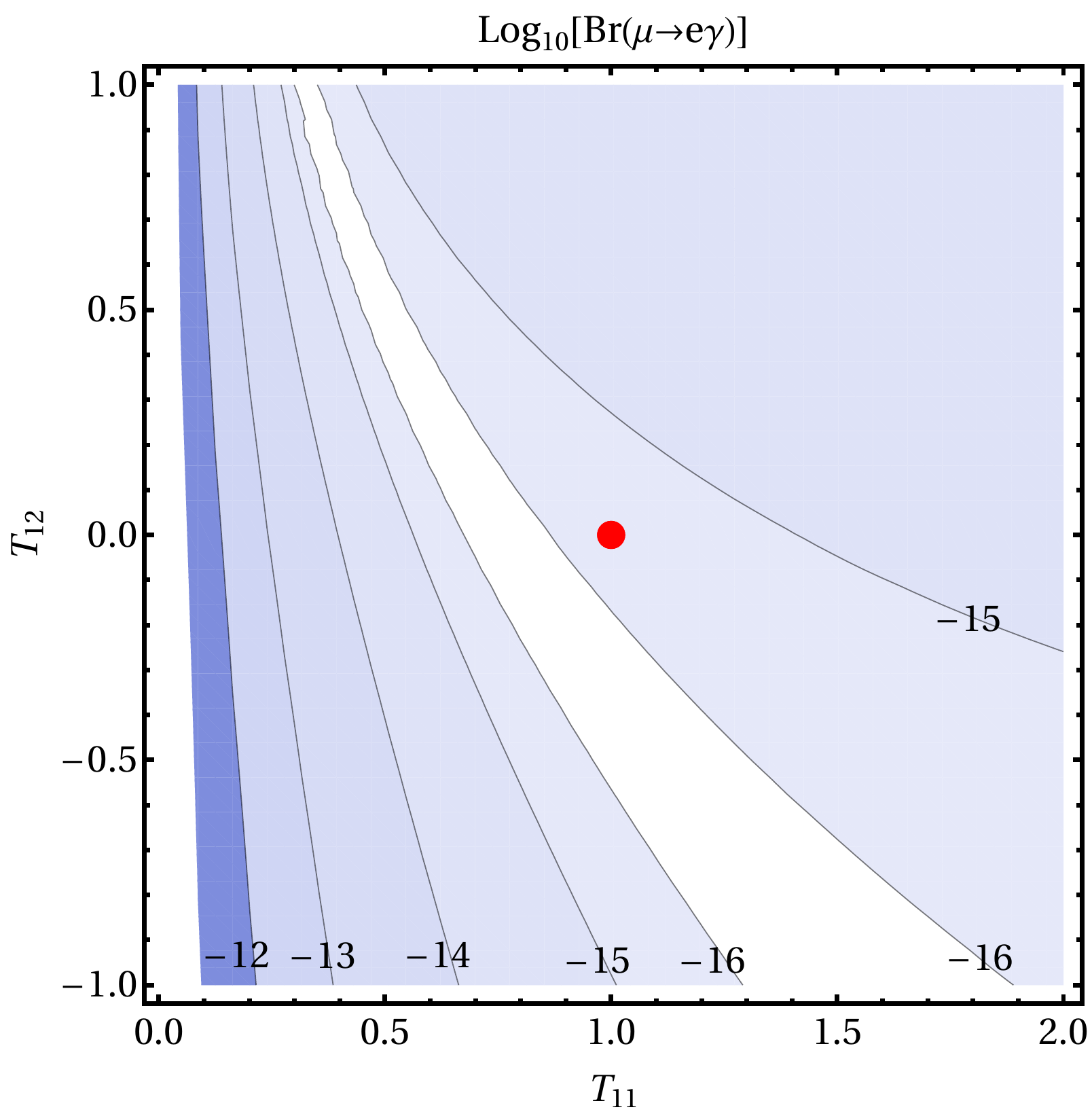}
 \caption{Example contours for Br$(\mu\to e\gamma)$ in the
   $(T_{11},T_{12})$ plane for the BNT model. Neutrino oscillation
   parameters are taken at their best fit points according to
   \cite{deSalas:2017kay}, while $M_{\Psi}=0.5$ TeV and $v_{\Phi}=10$
   KeV are fixed. All entries in the matrices $W$, $T$ and $K$ are
   taken to be \emph{trivial} ($W = T = \id$ and $K = 0$), except for
   $T_{11}$ and $T_{12}$, which are left as free parameters.  The
   point in the middle of the plot [$T_{11}=1,T_{12}=0$] corresponds
   to a simplified scenario with $T = \id$.
\label{fig:Ex2}}
\end{figure}

The application of the master parametrization is now straightforward. 
In the numerical scans that follow, the values of the neutrino 
oscillation parameters from the global fit \cite{deSalas:2017kay} are 
imposed, thus guaranteeing the consistency with oscillation experiments. 
We have implemented the model in {\tt SARAH} \cite{Staub:2013tta} and
obtained numerical results with the help of {\tt SPheno}
\cite{Porod:2011nf}. In the following we concentrate on the lepton
flavor violating (LFV) phenomenology of the model. The LFV observables
have been computed with {\tt FlavorKit} \cite{Porod:2014xia}. Some
selected results on the LFV observable Br$(\mu\to e\gamma)$ are shown
in Figs. \ref{fig:Ex1} and \ref{fig:Ex2}. When running a numerical
scan of the BNT model, one can assume specific simple forms for the
matrices that appear in the master parametrization (such as $T = \id$
or $K = 0$) or cover more general parameter
regions. Fig. \ref{fig:Ex1} shows the results of a random scan
with/without using the freedom in the matrices $T$ and $K$ as a
function of $v_{\Phi}$, while Fig. \ref{fig:Ex2} shows a contour plot
in the plane [$T_{11},T_{12}$]. Both examples serve to demonstrate
that it is important to scan over all allowed degrees of freedom in
order to obtain a general result. \\

{\em Final discussion:} The master parametrization allows one to
explore the parameter space of any Majorana neutrino mass model in a
complete way, while fixing at the same time the parameters to be in
agreement with all neutrino data.  The master parametrization is easy
to program, thus making parameter space exploration more direct than
ever. The master parametrization may also provide analytical insight
on some scenarios.

The application of the master parametrization is
straightforward. First, one must use the information from neutrino
oscillation experiments to fix the light neutrino masses and leptonic
mixing angles appearing in $\bar D_{\sqrt{m}}$ and $U$,
respectively. Then, by comparing the expression for the mass matrix of
the light neutrinos in the model under consideration with the general
master formula in Eq. \eqref{eq:master} one can easily identify the
global factor $f$, the Yukawa matrices $y_1$ and $y_2$ as well as the
matrix $M$. The latter can be singular-value decomposed to determine
$\Sigma$, $V_1$ and $V_2$, while the Yukawa matrices $y_1$ and $y_2$
are expressed in terms of a set of matrices ($\widehat W$,
$X_{1,2,3}$, $\bar B$, $T$, $K$ and $C_{1,2}$) by means of the master
parametrization in Eqs. \eqref{eq:par1} and \eqref{eq:par2}. In a
numerical analysis one can simply randomly scan over the free
parameters contained in these matrices to completely explore the
parameter space of a given model.

In closing, we should also point out some potential limitation of our
approach: In \textit{exceptional} cases, the master parametrization
may become either unnecessary, not direct or impractical. Exceptional
cases are simply those for which $y_1$ and $y_2$ are not completely
free parameters. A first category of exceptional models is given by
those with $y_1 = y_2 = \id$, such as in type-II seesaw. However, this
example of an unnecessary case can also trivially be solved.  More
involved situations are found in models with symmetric 
\cite{Anamiati:2016uxp} or antisymmetric 
\cite{Zee:1980ai,Cheng:1980qt,Zee:1985id,Babu:1988ki} Yukawa matrices, 
or models in which the Yukawa matrices have specific textures imposed 
by flavor symmetries. For such cases the master parametrization may be 
applicable only with additional constraints or become even impractical. 
We plan to return to a more detailed discussion of these cases in a 
future publication \cite{future}.

\section*{Acknowledgements}

The authors are grateful to Renato Fonseca and Claudia Hagedorn for
fruitful discussions. Work supported by the Spanish grants
AYA2015-66899-C2-1-P, SEV-2014-0398 and FPA2017-85216-P (AEI/FEDER, 
UE), PROMETEO/2018/165 and SEJI/2018/033 (Generalitat Valenciana) 
and the Spanish Red Consolider MultiDark FPA2017-90566-REDC.

\end{document}